\documentclass[a4paper,11pt]{article}

\usepackage[utf8]{inputenc}
\usepackage{hyperref,url}
\usepackage{enumitem}
\usepackage{amsmath}
\usepackage{multirow}
\usepackage{authblk}
\usepackage{graphicx}
\usepackage{times}
\usepackage{fullpage}

\hypersetup{
  colorlinks,
  linkcolor=black, 
  citecolor=black, 
  linktoc=all 
}

\usepackage{graphicx}
\usepackage{adjustbox} 
\usepackage{xcolor} 
\usepackage{textcomp} 
\AtBeginDocument{%
    \def\PYZsq{\textquotesingle}
}
\usepackage{upquote} 
\usepackage{eurosym} 
\usepackage{fancyvrb} 
\usepackage{enumitem}  
\usepackage{multirow}
\usepackage{subcaption}
\usepackage{placeins}

\makeatletter
\definecolor{urlcolor}{rgb}{0,.145,.698}
\definecolor{linkcolor}{rgb}{.71,0.21,0.01}
\definecolor{citecolor}{rgb}{.12,.54,.11}

\definecolor{ansi-black}{HTML}{3E424D}
\definecolor{ansi-black-intense}{HTML}{282C36}
\definecolor{ansi-red}{HTML}{E75C58}
\definecolor{ansi-red-intense}{HTML}{B22B31}
\definecolor{ansi-green}{HTML}{00A250}
\definecolor{ansi-green-intense}{HTML}{007427}
\definecolor{ansi-yellow}{HTML}{DDB62B}
\definecolor{ansi-yellow-intense}{HTML}{B27D12}
\definecolor{ansi-blue}{HTML}{208FFB}
\definecolor{ansi-blue-intense}{HTML}{0065CA}
\definecolor{ansi-magenta}{HTML}{D160C4}
\definecolor{ansi-magenta-intense}{HTML}{A03196}
\definecolor{ansi-cyan}{HTML}{60C6C8}
\definecolor{ansi-cyan-intense}{HTML}{258F8F}
\definecolor{ansi-white}{HTML}{C5C1B4}
\definecolor{ansi-white-intense}{HTML}{A1A6B2}
\definecolor{ansi-default-inverse-fg}{HTML}{FFFFFF}
\definecolor{ansi-default-inverse-bg}{HTML}{000000}

\DefineVerbatimEnvironment{Highlighting}{Verbatim}{commandchars=\\\{\}}

\def\PY@reset{\let\PY@it=\relax \let\PY@bf=\relax%
    \let\PY@ul=\relax \let\PY@tc=\relax%
    \let\PY@bc=\relax \let\PY@ff=\relax}
\def\PY@tok#1{\csname PY@tok@#1\endcsname}
\def\PY@toks#1+{\ifx\relax#1\empty\else%
    \PY@tok{#1}\expandafter\PY@toks\fi}
\def\PY@do#1{\PY@bc{\PY@tc{\PY@ul{%
    \PY@it{\PY@bf{\PY@ff{#1}}}}}}}
\def\PY#1#2{\PY@reset\PY@toks#1+\relax+\PY@do{#2}}

\expandafter\def\csname PY@tok@w\endcsname{\def\PY@tc##1{\textcolor[rgb]{0.73,0.73,0.73}{##1}}}
\expandafter\def\csname PY@tok@c\endcsname{\let\PY@it=\textit\def\PY@tc##1{\textcolor[rgb]{0.25,0.50,0.50}{##1}}}
\expandafter\def\csname PY@tok@cp\endcsname{\def\PY@tc##1{\textcolor[rgb]{0.74,0.48,0.00}{##1}}}
\expandafter\def\csname PY@tok@k\endcsname{\let\PY@bf=\textbf\def\PY@tc##1{\textcolor[rgb]{0.00,0.50,0.00}{##1}}}
\expandafter\def\csname PY@tok@kp\endcsname{\def\PY@tc##1{\textcolor[rgb]{0.00,0.50,0.00}{##1}}}
\expandafter\def\csname PY@tok@kt\endcsname{\def\PY@tc##1{\textcolor[rgb]{0.69,0.00,0.25}{##1}}}
\expandafter\def\csname PY@tok@o\endcsname{\def\PY@tc##1{\textcolor[rgb]{0.40,0.40,0.40}{##1}}}
\expandafter\def\csname PY@tok@ow\endcsname{\let\PY@bf=\textbf\def\PY@tc##1{\textcolor[rgb]{0.67,0.13,1.00}{##1}}}
\expandafter\def\csname PY@tok@nb\endcsname{\def\PY@tc##1{\textcolor[rgb]{0.00,0.50,0.00}{##1}}}
\expandafter\def\csname PY@tok@nf\endcsname{\def\PY@tc##1{\textcolor[rgb]{0.00,0.00,1.00}{##1}}}
\expandafter\def\csname PY@tok@nc\endcsname{\let\PY@bf=\textbf\def\PY@tc##1{\textcolor[rgb]{0.00,0.00,1.00}{##1}}}
\expandafter\def\csname PY@tok@nn\endcsname{\let\PY@bf=\textbf\def\PY@tc##1{\textcolor[rgb]{0.00,0.00,1.00}{##1}}}
\expandafter\def\csname PY@tok@ne\endcsname{\let\PY@bf=\textbf\def\PY@tc##1{\textcolor[rgb]{0.82,0.25,0.23}{##1}}}
\expandafter\def\csname PY@tok@nv\endcsname{\def\PY@tc##1{\textcolor[rgb]{0.10,0.09,0.49}{##1}}}
\expandafter\def\csname PY@tok@no\endcsname{\def\PY@tc##1{\textcolor[rgb]{0.53,0.00,0.00}{##1}}}
\expandafter\def\csname PY@tok@nl\endcsname{\def\PY@tc##1{\textcolor[rgb]{0.63,0.63,0.00}{##1}}}
\expandafter\def\csname PY@tok@ni\endcsname{\let\PY@bf=\textbf\def\PY@tc##1{\textcolor[rgb]{0.60,0.60,0.60}{##1}}}
\expandafter\def\csname PY@tok@na\endcsname{\def\PY@tc##1{\textcolor[rgb]{0.49,0.56,0.16}{##1}}}
\expandafter\def\csname PY@tok@nt\endcsname{\let\PY@bf=\textbf\def\PY@tc##1{\textcolor[rgb]{0.00,0.50,0.00}{##1}}}
\expandafter\def\csname PY@tok@nd\endcsname{\def\PY@tc##1{\textcolor[rgb]{0.67,0.13,1.00}{##1}}}
\expandafter\def\csname PY@tok@s\endcsname{\def\PY@tc##1{\textcolor[rgb]{0.73,0.13,0.13}{##1}}}
\expandafter\def\csname PY@tok@sd\endcsname{\let\PY@it=\textit\def\PY@tc##1{\textcolor[rgb]{0.73,0.13,0.13}{##1}}}
\expandafter\def\csname PY@tok@si\endcsname{\let\PY@bf=\textbf\def\PY@tc##1{\textcolor[rgb]{0.73,0.40,0.53}{##1}}}
\expandafter\def\csname PY@tok@se\endcsname{\let\PY@bf=\textbf\def\PY@tc##1{\textcolor[rgb]{0.73,0.40,0.13}{##1}}}
\expandafter\def\csname PY@tok@sr\endcsname{\def\PY@tc##1{\textcolor[rgb]{0.73,0.40,0.53}{##1}}}
\expandafter\def\csname PY@tok@ss\endcsname{\def\PY@tc##1{\textcolor[rgb]{0.10,0.09,0.49}{##1}}}
\expandafter\def\csname PY@tok@sx\endcsname{\def\PY@tc##1{\textcolor[rgb]{0.00,0.50,0.00}{##1}}}
\expandafter\def\csname PY@tok@m\endcsname{\def\PY@tc##1{\textcolor[rgb]{0.40,0.40,0.40}{##1}}}
\expandafter\def\csname PY@tok@gh\endcsname{\let\PY@bf=\textbf\def\PY@tc##1{\textcolor[rgb]{0.00,0.00,0.50}{##1}}}
\expandafter\def\csname PY@tok@gu\endcsname{\let\PY@bf=\textbf\def\PY@tc##1{\textcolor[rgb]{0.50,0.00,0.50}{##1}}}
\expandafter\def\csname PY@tok@gd\endcsname{\def\PY@tc##1{\textcolor[rgb]{0.63,0.00,0.00}{##1}}}
\expandafter\def\csname PY@tok@gi\endcsname{\def\PY@tc##1{\textcolor[rgb]{0.00,0.63,0.00}{##1}}}
\expandafter\def\csname PY@tok@gr\endcsname{\def\PY@tc##1{\textcolor[rgb]{1.00,0.00,0.00}{##1}}}
\expandafter\def\csname PY@tok@ge\endcsname{\let\PY@it=\textit}
\expandafter\def\csname PY@tok@gs\endcsname{\let\PY@bf=\textbf}
\expandafter\def\csname PY@tok@gp\endcsname{\let\PY@bf=\textbf\def\PY@tc##1{\textcolor[rgb]{0.00,0.00,0.50}{##1}}}
\expandafter\def\csname PY@tok@go\endcsname{\def\PY@tc##1{\textcolor[rgb]{0.53,0.53,0.53}{##1}}}
\expandafter\def\csname PY@tok@gt\endcsname{\def\PY@tc##1{\textcolor[rgb]{0.00,0.27,0.87}{##1}}}
\expandafter\def\csname PY@tok@err\endcsname{\def\PY@bc##1{\setlength{\fboxsep}{0pt}\fcolorbox[rgb]{1.00,0.00,0.00}{1,1,1}{\strut ##1}}}
\expandafter\def\csname PY@tok@kc\endcsname{\let\PY@bf=\textbf\def\PY@tc##1{\textcolor[rgb]{0.00,0.50,0.00}{##1}}}
\expandafter\def\csname PY@tok@kd\endcsname{\let\PY@bf=\textbf\def\PY@tc##1{\textcolor[rgb]{0.00,0.50,0.00}{##1}}}
\expandafter\def\csname PY@tok@kn\endcsname{\let\PY@bf=\textbf\def\PY@tc##1{\textcolor[rgb]{0.00,0.50,0.00}{##1}}}
\expandafter\def\csname PY@tok@kr\endcsname{\let\PY@bf=\textbf\def\PY@tc##1{\textcolor[rgb]{0.00,0.50,0.00}{##1}}}
\expandafter\def\csname PY@tok@bp\endcsname{\def\PY@tc##1{\textcolor[rgb]{0.00,0.50,0.00}{##1}}}
\expandafter\def\csname PY@tok@fm\endcsname{\def\PY@tc##1{\textcolor[rgb]{0.00,0.00,1.00}{##1}}}
\expandafter\def\csname PY@tok@vc\endcsname{\def\PY@tc##1{\textcolor[rgb]{0.10,0.09,0.49}{##1}}}
\expandafter\def\csname PY@tok@vg\endcsname{\def\PY@tc##1{\textcolor[rgb]{0.10,0.09,0.49}{##1}}}
\expandafter\def\csname PY@tok@vi\endcsname{\def\PY@tc##1{\textcolor[rgb]{0.10,0.09,0.49}{##1}}}
\expandafter\def\csname PY@tok@vm\endcsname{\def\PY@tc##1{\textcolor[rgb]{0.10,0.09,0.49}{##1}}}
\expandafter\def\csname PY@tok@sa\endcsname{\def\PY@tc##1{\textcolor[rgb]{0.73,0.13,0.13}{##1}}}
\expandafter\def\csname PY@tok@sb\endcsname{\def\PY@tc##1{\textcolor[rgb]{0.73,0.13,0.13}{##1}}}
\expandafter\def\csname PY@tok@sc\endcsname{\def\PY@tc##1{\textcolor[rgb]{0.73,0.13,0.13}{##1}}}
\expandafter\def\csname PY@tok@dl\endcsname{\def\PY@tc##1{\textcolor[rgb]{0.73,0.13,0.13}{##1}}}
\expandafter\def\csname PY@tok@s2\endcsname{\def\PY@tc##1{\textcolor[rgb]{0.73,0.13,0.13}{##1}}}
\expandafter\def\csname PY@tok@sh\endcsname{\def\PY@tc##1{\textcolor[rgb]{0.73,0.13,0.13}{##1}}}
\expandafter\def\csname PY@tok@s1\endcsname{\def\PY@tc##1{\textcolor[rgb]{0.73,0.13,0.13}{##1}}}
\expandafter\def\csname PY@tok@mb\endcsname{\def\PY@tc##1{\textcolor[rgb]{0.40,0.40,0.40}{##1}}}
\expandafter\def\csname PY@tok@mf\endcsname{\def\PY@tc##1{\textcolor[rgb]{0.40,0.40,0.40}{##1}}}
\expandafter\def\csname PY@tok@mh\endcsname{\def\PY@tc##1{\textcolor[rgb]{0.40,0.40,0.40}{##1}}}
\expandafter\def\csname PY@tok@mi\endcsname{\def\PY@tc##1{\textcolor[rgb]{0.40,0.40,0.40}{##1}}}
\expandafter\def\csname PY@tok@il\endcsname{\def\PY@tc##1{\textcolor[rgb]{0.40,0.40,0.40}{##1}}}
\expandafter\def\csname PY@tok@mo\endcsname{\def\PY@tc##1{\textcolor[rgb]{0.40,0.40,0.40}{##1}}}
\expandafter\def\csname PY@tok@ch\endcsname{\let\PY@it=\textit\def\PY@tc##1{\textcolor[rgb]{0.25,0.50,0.50}{##1}}}
\expandafter\def\csname PY@tok@cm\endcsname{\let\PY@it=\textit\def\PY@tc##1{\textcolor[rgb]{0.25,0.50,0.50}{##1}}}
\expandafter\def\csname PY@tok@cpf\endcsname{\let\PY@it=\textit\def\PY@tc##1{\textcolor[rgb]{0.25,0.50,0.50}{##1}}}
\expandafter\def\csname PY@tok@c1\endcsname{\let\PY@it=\textit\def\PY@tc##1{\textcolor[rgb]{0.25,0.50,0.50}{##1}}}
\expandafter\def\csname PY@tok@cs\endcsname{\let\PY@it=\textit\def\PY@tc##1{\textcolor[rgb]{0.25,0.50,0.50}{##1}}}

\def\PYZus{\char`\_}

\def\PYZsh{\char`\#}
\def\PYZpc{\char`\%}

\def\PYZsq{\char`\'}

\definecolor{incolor}{rgb}{0.0, 0.0, 0.5}
\definecolor{outcolor}{rgb}{0.545, 0.0, 0.0}
\makeatother

\title{SCALPEL3: a scalable open-source library for healthcare claims databases}

\author[1,2]{Emmanuel Bacry}
\author[3]{St\'ephane Ga\"iffas}
\author[4]{Fanny Leroy}
\author[2]{Maryan Morel}
\author[2,4]{Dinh Phong Nguyen}
\author[2]{Youcef Sebiat}
\author[2]{Dian Sun}

\affil[1]{CEREMADE, Université Paris-Dauphine, PSL, Paris, France}
\affil[3]{LPSM, Université de Paris, Paris, France}
\affil[2]{CMAP, Ecole polytechnique, Palaiseau, France}
\affil[4]{Caisse Nationale de l’Assurance Maladie, 75986 Paris Cedex 20, France}

\begin{document}

\maketitle



\begin{abstract}
\noindent
\emph{Objective\textup:} This article introduces SCALPEL3 (SCAlable Pipeline for hEaLth data), a scalable open-source framework for studies involving Large Observational Databases (LODs). It focuses on scalable medical concept extraction, easy interactive analysis, and helpers for data flow analysis to accelerate studies performed on LODs.

\noindent
\emph{Materials and methods\textup:} Inspired from web analytics, SCALPEL3 rely on distributed computing, data denormalization and columnar storage. It was compared to the existing SAS-Oracle SNDS infrastructure by performing several queries on a dataset containing a three years-long history of healthcare claims of 13.7 million patients.

\noindent
\emph{Results and Discussion\textup:} SCALPEL3 horizontal scalability allows handling large tasks quicker than the existing infrastructure while it has comparable performance when using only a few executors.
SCALPEL3 provides a sharp interactive control of data processing through legible code, which helps to build studies with full reproducibility, leading to improved maintainability and audit of studies performed on LODs.

\noindent
\emph{Conclusion\textup:} SCALPEL3 makes studies based on SNDS much easier and more scalable than the existing framework~\cite{Tuppin:2017}. 
It is now used at the agency collecting SNDS data, at the French Ministry of Health and soon at the National Health Data Hub in France~\cite{HealthDataHub}.

\end{abstract}


\section{Introduction}
\label{sec:introduction}

In the past decade, the volume of healthcare data and its accessibility rose quickly.
For instance, in France, the SNDS claims database contained 86\% of the French population in 2010~\cite{Tuppin:2010} to reach 98.8\% in 2015~\cite{Tuppin:2017} leading to one of the world's largest health Large Observational Database (LOD)~\cite{Tuppin:2017,Bezin2017}. 
The exhaustivity of LODs such as SNDS has proven useful for public health research, by improving the statistical power of algorithms using this data and by mitigating the sensitivity to selection biases~\cite{Tuppin:2017}.

However, such an abundance of data comes at a cost: SNDS is a very complex database, with data spread across hundreds of tables and columns.
Its scale makes data manipulation non-trivial.
More importantly, using this data requires a tremendous amount of knowledge from SNDS experts. Many coding or data recording subtleties, such as data duplication caused by administrative complexity, might bewilder inexperienced users.
Deriving proper health events definitions and extracting them accurately is, therefore, a difficult task, having important consequences on the derived studies~\cite{Tuppin:2017, Hansen2013}.
These issues are of course not unique to SNDS but shared by many LODs~\cite{Madigan2014a}.

This paper proposes an answer to this problem by introducing SCALPEL3 (SCALable Pipeline for hEaLth data), an open-source framework intending to reduce such entry barriers to LODs.
This framework attempts to simplify medical concept extraction by providing a set of tools performing batch Extract-Transform-Load (ETL) tasks, while an interactive API eases the manipulation and the exploration of longitudinal cohorts.
Thus, this research focuses on the following objectives:
\begin{enumerate}
    \item Design and implement a scalable tool allowing to extract and manipulate longitudinal patient data from large observational databases;
    \item Simplify methodological research by reducing SNDS data complexity and by easing data loading into formats used by common machine learning libraries;
    \item Foster reproducibility by monitoring the data flow and by following best practices for clean code;
    \item Promote reusability and extensibility by documenting and open-sourcing SCALPEL3 implementation.
\end{enumerate}

The main concepts used by SCALPEL3 and some related works are presented in Section~\ref{sec:bg}. 
The LOD for which SCALPEL3 was initially designed for is described in Section~\ref{sec:matmet}, together with SCALPEL3 methods and abstractions.
The scalability of SCALPEL3 is evaluated in Section~\ref{sec:results}, while Section~\ref{sec:discussion} discusses its strengths and limitations.

\section{Background}
\label{sec:bg}

LODs are not designed to perform medical research. Electronic Health Records (EHR) data directly supports clinical care and are used to justify care billing and reimbursement, while claims data are primarily used for reimbursement purpose. The data models and terminologies used in such databases were optimized to suit these particular goals, resulting in normalized data models built around hospital stays, transactions, or cash flows~\cite{Tuppin:2017}.
Extracting meaningful patients care pathways from such data can be decomposed into two tasks. 
First, all the data corresponding to a set of patients need to be identified and collected. 
When the data is not normalized around the patients, this task requires several join operations which can be very costly in terms of computations as the data volume increases.
Second, medical concepts have to be properly identified from administrative codes: this \emph{phenotyping} task relies heavily on a combination of medical and database knowledge.
The algorithms used to perform concept extraction from administrative data are either disclosed through scientific publications or shared as lengthy SQL queries~\cite{looten2019studies}. 
Their code or the description of the algorithms involved can vary in quality, hindering reuse, and reproducibility. As a result, building a study from scratch might be faster than reusing poorly documented code from previous works~\cite{peng2006reproducible,looten2019studies}.
Besides, access to LODs such as SNDS might rely on proprietary software such as SAS~\cite{sas} or SPSS~\cite{spss}. 
While these tools are suitable to produce public health studies, they hinder methodological research as they do not interact easily with R or Python packages that implement state-of-the-art machine learning algorithms.
All of these challenges are complex to solve and exacerbated by the data volume at hand.

\subsection*{Related works}

Several research programs produce tools in order to alleviate some of these issues. 
An important research effort aims at easing data integration and interoperability by producing standard data models and terminologies to be shared across institutions.
Observational Medical Outcomes Partnership Common Data Model (OMOP CDM), which is supported by the Observational Health Data Sciences and Informatics (OHDSI) research program~\cite{hripcsak2015observational}, and the Informatics for Integrating Biology \& the Bedside (i2b2) data model~\cite{murphy2010serving}, can be considered as the most pervasive data models developed for this purpose. 
OMOP CDM can be used to standardize EHR or claims data, while i2b2 is focused on EHR data. 

Both models are centered around the patients, thus reducing the number of join operations required to access a specific patient history. They also rely on a normalized data model combined with SQL databases.
A collection of open-source software has been developed on top of these models, implementing analytics or visualization tools~\cite{huser2016multisite}. These softwares can take the form of R libraries~\cite{huser2016multisite}, or compiled Java~\cite{WhiteRabbit} programs with a graphical user interface. While making these softwares freely available is an important step to foster methodological research, they do not seem to be easily extensible or interoperable as they do not provide documented APIs to build new software upon it.
Besides, the process of transforming an existing database in order to conform to such standards is costly, as it requires to build complex mappings between shared representations expressed through highly heterogeneous codes from one information system to the other.
In the case of the SNDS database considered in this work, such a mapping is still work in progress~\cite{doutreligne2020alignement}.

In other fields, web-scale analytics have shifted from the use of normalized SQL databases towards NoSQL technologies relying on distributed computing, denormalization, and columnar storage. The use of distributed computing allowed gains in computational power using low cost, commodity servers instead of expensive dedicated hardware~\cite{bonner2017exploring}.
A work from OHDSI~\cite{powers2016apache} compared the ACHILLES software (R~\cite{R}, PostgreSQL~\cite{postgresql1996postgresql}) with Apache Spark~\cite{Zaharia:2016} using common SQL requests.
They observed performance gains for Spark even on a single server or small clusters, at the exception of requests leading to large network I/O, since such operations are known to be the slowest operations in a distributed computing framework because of network latencies and throughput.
It can create bottlenecks when many data chunks are sent across the servers in the cluster to perform a join or a groupby operation (leading to so-called \emph{shuffles}).
Denormalization can be a way to circumvent this issue by performing a set of join operations beforehand, once and for all~\cite{wei2008service,li2014widetable,dehdouh2015using}, reducing join operations to simple look-ups over a very large table. 
The data duplication resulting from such joins operations might lead to storage issues, which can be mitigated with the help of columnar storage formats~\cite{li2014widetable,Melnik:2010} using compression strategies.

To the best of our knowledge, such an approach has not been implemented to perform ETL on large health databases. Prior works are either relying on SQL and normalized schemas~\cite{jannot2017georges,ong2017dynamic} or applied to small datasets~\cite{Harris2018a}.
This paper describes and implements such an approach for large health databases, as explained in the next section.

\section{Material and Methods}
\label{sec:matmet}

This work focuses on (i) denormalizing the data in combination with columnar storage and distributed computing to perform concept extraction, (ii) providing a structured and re-usable concept library, and (iii) introduce useful abstractions to handle cohort data. 
Scalability issues are handled by (i), while (ii) and (iii) foster the reuse of code and knowledge across studies.
This is achieved by reducing both study-specific code and database entry barriers by providing ready-to-use concepts. 
SCALPEL3 provides Scala~\cite{odersky2004overview} and Python APIs to ensure easy extension and interoperability with numerous libraries. 
All the code supporting this paper is open source and freely available. 

This paper is not about data integration from disparate sources, such as multiple EHR systems, but rather about an ETL based on batch distributed processing of a large, centralized claims database.

\subsection{The SNDS database}
\label{subsec:snds_description}

This work was performed using the \emph{Système National des Données de Santé} (SNDS), a large claims database containing pseudonymized data on 98.8\% of the French population (66~million patients in 2015) \cite{Tuppin:2017,Bezin2017}. 
It contains time-stamped information about medical events leading to reimbursement (see Table 1 in \cite{Tuppin:2017} for an exhaustive list of available data) in the last 3 years\footnote{which can be extended up to 20 years under some restrictions.}.
It contains more than 20~billion health events per year, representing roughly 70TB of data.

SNDS is composed of multiple ``sub-databases'', each one with a star schema.
The central table records events leading to cash flows that need to be joined to many other tables to access medical information\footnote{We work with two main sub-databases containing data relevant for public-health research. When working on drug safety studies, each of these two databases contains 8 relevant tables, representing approximately 5~billion lines per year when restricted to $65+$ y.o. subjects.}.\
In this form,  retrieving patient information for statistical studies is very costly in terms of computation and expert knowledge: targeted data can be spread across multiple databases, tens of tables, and hundreds of columns, and its identification requires a deep administrative knowledge of the French health-care reimbursement mechanisms.
Mitigating these issues is precisely the motivation of the SCALPEL3 framework.

\subsection{SCALPEL3\textup: a SCAlable Pipeline for hEaLth data}
\label{subsec:pipeline}


SCALPEL3 is based on Apache Spark~\cite{Zaharia:2016}, a robust and widely adopted distributed in-memory computation framework. Spark provides a powerful SQL-like high-level API and a more granular API to perform data operations. It can be coupled with the Hadoop File System (HDFS)~\cite{Shvachko:2010:HDF} replication system to accelerate large files reading and distribution over a computing cluster. SCALPEL3 is an open-source framework organized in the following three components.

\medskip
\noindent
\textbf{SCALPEL-Flattening~\cite{SCALPEL-Flattening}} denormalizes the data ``once and for all'' to avoid joining many tables each time the data of a patient is accessed.
Its input is a set of CSV files extracted from the original SNDS database.

\medskip
\noindent
\textbf{SCALPEL-Extraction~\cite{SCALPEL-Extraction}} defines concepts extractors that process the denormalized data and transformers, that compute more complex events based on extractors output. For example, extractors can fetch all drug dispenses or medical acts.

\medskip
\noindent
\textbf{SCALPEL-Analysis~\cite{SCALPEL-Analysis}} implements powerful and scalable abstractions that can be used for data analysis, such as easy ways to investigate data quality issues.
It can load data into formats commonly used in machine learning, such as TensorFlow or PyTorch tensors or NumPy arrays.

\begin{figure}[!hp]
    \centering
    \includegraphics[width=0.85\textwidth]{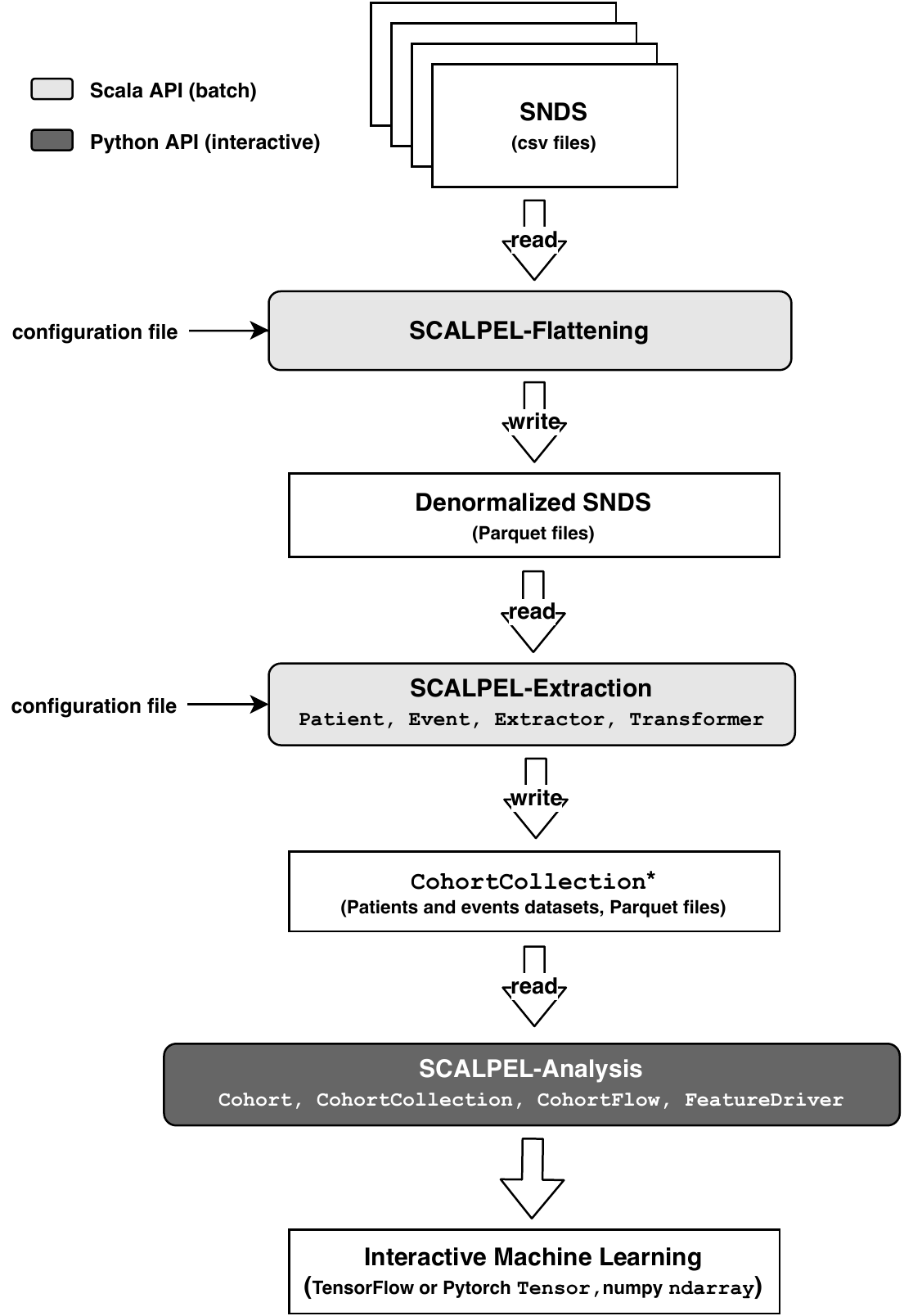}
    \caption{SCALPEL3 workflow. SCALPEL3 is made of three independent open-source libraries plugged one after another. SCALPEL-Flattening, which is implemented in Scala/Spark, denormalizes the input database exported as CSV or Parquet files into a single big flat database. Then, SCALPEL-Extraction, implemented in Scala/Spark, extracts concepts from this flat database. Finally, SCALPEL-Analysis, implemented in Python/PySpark loads extracted concepts to perform in-memory interactive analysis and feed machine learning algorithms.}
    \label{fig:workflow}
\end{figure}

\medskip
As SCALPEL-Flattening and SCALPEL-Extraction perform batch operations, they need to read (resp. write) input (resp. output) data from the file-system (local or HDFS).
They are implemented in Scala in order to access Spark's low-level API and take advantage of functional programming and static typing, resulting in rigorous automated testing (94\% of the Scala code is covered by unit tests).
Both can be configured through textual configuration files or be used as libraries.
SCALPEL-Analysis is a python module implemented in Python/PySpark and designed for interactive use. 
It can be used in a Jupyter notebook~\cite{Kluyver:2016aa} for instance. This workflow is illustrated in Figure~\ref{fig:workflow}.

\subsection{SCALPEL-Flattening\textup: denormalization of the data}
\label{subsubsec:flattening}

As mentioned earlier, performing data analysis on SNDS patients' health requires many joins and can consequently be extremely slow. To circumvent this issue, the data are denormalized by joining the tables sequentially to obtain a big table in which each line corresponds to a patient identifier and a wide representation of an event.

Denormalizing a star-schema database results in a really big table due to values replications.  To circumvent storage and computation issues, the denormalized data is stored in Parquet~\cite{Apache-Parquet} files, an open-source columnar storage format implementing Google's Dremel~\cite{Melnik:2010} data model. Parquet is well-integrated in the Spark ecosystem~\cite{Armbrust:2015}, allowing us to take advantage of the columnar storage in terms of data compression and query optimization.
SCALPEL-Flattening first converts the input CSV files containing exports of SNDS tables to Parquet files. 
Then, it recursively performs left joins with these tables, starting with the central table. Finally, it writes the results in a single Parquet file.
To ensure the scalability of these big join operations, the input data can be automatically divided with respect to some time unit (such as years, months) before performing the join operations. In this case, the joins results are sequentially appended to the output parquet file. These operations are repeated for each SNDS sub-databases.
The size of the temporal slicing used in the joins, the schema, and the joining keys can be tuned by the end-user through a configuration file, which defaults to the denormalization of tables containing only medical data (as opposed to econometric and administrative data). 
A set of statistics that monitors the denormalization process is automatically computed along the steps involved in it, in order to ensure that no loss of information occurs.

\subsection{SCALPEL-Extraction\textup: extraction of concepts}
\label{subsubsec:extraction}

SCALPEL-Extraction provides fast extractions of medical concepts from the denormalized tables produced by SCALPEL-Flattening.
By providing ready-to-use medical events, SCALPEL-extraction encapsulates SNDS technical knowledge but keeps medical data as raw as possible, so that end-users have access to fine-grained data which is critical when designing observational studies~\cite{Wang2016, Hong2018}.
The extracted concepts are organized around two abstractions: \texttt{Patient} and \texttt{Event}.

\medskip
\noindent
\textbf{The \texttt{Patient} abstraction} has a unique \texttt{patientID}, a \texttt{gender}, a \texttt{birthDate} and eventually a \texttt{deathDate}. 

\medskip
\noindent
\textbf{The \texttt{Event} abstraction} allows to represent any event associated to a patient.
It can be punctual (e.g., medical act) or continuous (e.g., hospitalization).

\medskip
All concepts are automatically extracted into \texttt{Patient} or \texttt{Event} objects by a set of \texttt{Extractor}s and \texttt{Transformer}s, designed to fetch the data in the relevant tables and columns of the SNDS \texttt{Source}s.

\begin{figure}[!hp]
    \centering
    \includegraphics[width=\textwidth]{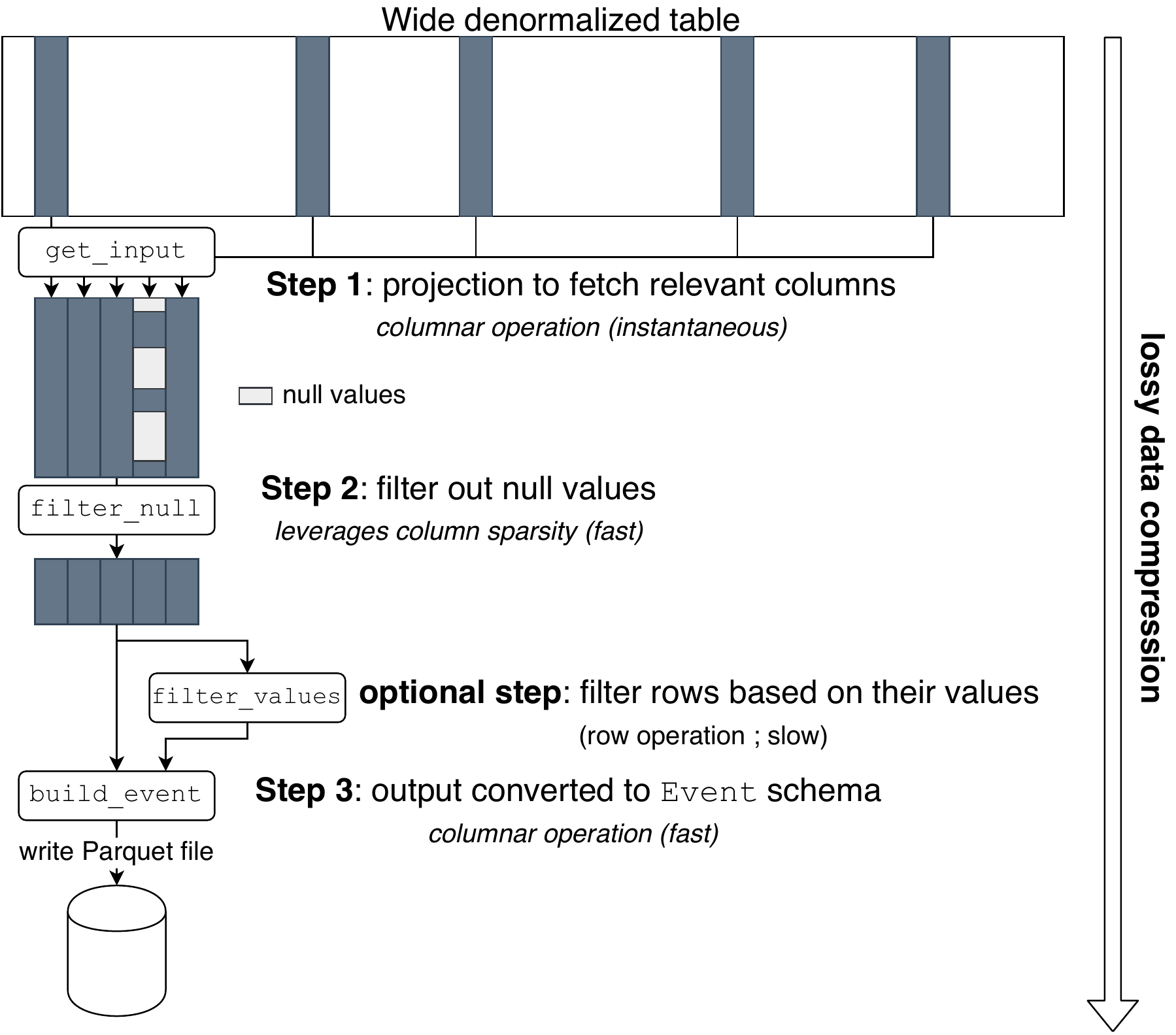}
    \caption{\texttt{Extractor} design. \texttt{Extractor}s implemented in SCALPEL-Extraction successively refines the input table (a large denormalized table) by taking advantage of fast columnar operations to produce ready-to-use medical events. Step~1 selects the relevant columns (equivalent to a hash table look-up) while Step~2 removes rows where null values are detected in specific columns, taking advantage of the sparsity of columnar representation (null values are not encoded in the data).
    Optionally, this extraction process filters out rows based on their values.
    Finally, Step~3 conforms the data to the \texttt{Event} schema, and is written to a Parquet file.}
    \label{fig:extractor}
\end{figure}

\medskip
\noindent
\textbf{The \texttt{Extractor} abstraction} maps a \texttt{Row} of a \texttt{Source} to zero or many \texttt{Event}s:
\begin{center}
    \texttt{Extractor}: \texttt{Row} $\mapsto$ \texttt{List[Event]}.
\end{center}
\texttt{Extractor}s successively refines data from the input (wide denormalized tables) by (1) identifying the relevant columns, (2) filtering out null values according to some columns and (3) conform the extracted data to a standardized schema. These three operations are very fast when performed on columnar data, as they exploit sparsity (null values are not represented in the data) and consist in simple look-ups over hash tables containing columns metadata. 
An optional step that filters rows by value can occur before step (3). 
This operation is slower as it manipulates row values, but since it is performed near the end of the extraction process, it typically occurs on small data. 
This process is illustrated Figure~\ref{fig:extractor}.

Many extractors are available to fetch medical acts, diagnoses, hospital stays, among others, an example being the drug dispense \texttt{Extractor} which allows extracting events related to specific subsets of drugs and to output events at multiple levels of granularity (drug, molecule, ATC class, custom classes) as defined in a configuration file. 
This simple architecture makes it easy to add new \texttt{Extractor}s and to answer to any extraction need.

\medskip
\noindent
\textbf{The \texttt{Transformer} abstraction} transforms a collection of \texttt{Event}s related to a unique \texttt{Patient} into a list of more complex \texttt{Event}s (complex diseases, drug exposures, \ldots): 
\begin{center}
    \texttt{Transformer}: \texttt{List[Event]} $\mapsto$ \texttt{List[Event]}.
\end{center}

A \texttt{Transformer} is based on specific algorithms requiring multidisciplinary knowledge from epidemiologists, statisticians, clinicians, physicians, and SNDS experts~\cite{Tuppin:2017}. \texttt{Transformer}s usually combine events built by \texttt{Extractor}s to build more complex events, such as computing drug exposures from timestamped drug dispenses.
\texttt{Extractor}s and \texttt{Transformer}s can be used through a Scala API or controlled using a textual configuration file.
Many \texttt{Transformer}s used in several studies such as~\cite{morel2017convsccs,neumann2012pioglitazone} are implemented and ready to use.  

Besides Parquet files containing extracted events, SCALPEL-Extraction outputs metadata tracking the data used to build each type of extracted events. This file can be leveraged by SCALPEL-Analysis to build \texttt{Cohort}s and flowcharts, as explained below.

\subsection{SCALPEL-Analysis\textup: interactive manipulation and analysis of cohorts}
\label{subsubsec:exploration}

While SCALPEL-Flattening and SCALPEL-Extraction are implemented in Scala/Spark for performance and maintainability, SCALPEL-Analysis is implemented in Python/PySpark~\cite{Zaharia:2016} since it is designed for interactive environments, such as Jupyter notebooks~\cite{Kluyver:2016aa}. SCALPEL-Analysis eases the manipulation and analysis of cohort data. 
It is based on the following abstractions:

\medskip
\noindent
\textbf{The \texttt{Cohort} abstraction} is a set of \texttt{Patient}s and their associated \texttt{Event}s in a [\texttt{startDate}, \texttt{endDate}] time-window.
Basic operations such as union, intersection, and difference can be performed between \texttt{Cohort}s, while a human-readable description is automatically updated in the results.
More granular control is kept available through accesses to the underlying Spark DataFrames (using Spark DataFrame API).
This combination allows easy data engineering and fine-grained, yet reproducible, experiments.

\medskip
\noindent
\textbf{The \texttt{CohortCollection} abstraction} is a collection of \texttt{Cohort}s on which operations can be jointly performed. The \texttt{CohortCollection} has metadata that keeps the information about each \texttt{Cohort}, such as
the successive operations performed on it, the Parquet files they are stored in and a git commit hash of the code producing the extraction from the \texttt{Source}.

\medskip
International guidelines~\cite{benchimol2015reporting} regarding studies based on LODs insist on the explanation of cohort construction to highlight eventual population biases, motivating the following \texttt{CohortFlow} abstraction.

\medskip
\noindent
\textbf{The \texttt{CohortFlow} abstraction} is an ordered iterator defined as the following left fold operation
\begin{equation*}
    \text{foldl}(c: \text{\texttt{CohortCollection}}, \cap) := (((c_0 \cap c_1) \cap c_2) \cap \dots c_n )
\end{equation*}
assuming an input \texttt{CohortCollection} $c$ of length $n$, where $\cap$ denotes an intersection of the \texttt{Cohort}s' patients.
It is meant to track the stages leading to a final \texttt{Cohort}, where each intermediate \texttt{Cohort} is stored along with textual information about the filtering rules used to go from each stage to the next one.

\medskip
\noindent
\textbf{The \texttt{scalpel.stats} module} produces descriptive statistics on a \texttt{Cohort} and their associated plots. 
For now, it contains more than 25 \texttt{Patient}-centric or \texttt{Event}-centric statistics, adding a custom one being very easy.
Among other things, this module provides automatic reporting as text or graphical displays, with performance optimization through data caching.
It can be combined with \texttt{CohortFlow} to compute various statistics at each analysis stage, to assess the biases induced along with successive population filtering operations.
Flowcharts can easily be produced to track how many subjects were removed at each stage. Flowcharts can be produced either from a \texttt{CohortFlow}, or the metadata tracking the data extraction process produced by SCALPEL-Extraction.
Examples are provided in Supplementary Material.

\medskip
SCALPEL-Analysis also provides tools producing datasets in formats compatible with popular machine learning libraries.
At the core of these tools is the \texttt{FeatureDriver} abstraction.

\medskip
\noindent
\textbf{The \texttt{FeatureDriver} abstraction} is used to transform \texttt{Cohort}s into data formats suitable for machine learning algorithms, such as \texttt{numpy.ndarray}~\cite{numpy}, \texttt{tensorflow.tensor}~\cite{tensorflow2015-whitepaper} and \texttt{pytorch.tensor}~\cite{paszke2017automatic}.
It is mainly a transformation of a Spark dataframe representation into a tensor-based format.
\texttt{FeatureDriver}s perform several sanity checks, such as time-zone and event dates consistency, and can be easily extended by end-users, thanks to the PySpark API.

\section{Results}
\label{sec:results}

Scaling experiments presented in this section were performed on a SNDS subset containing 13.7~million patients followed up to three years described in Table~\ref{tab:dataset}. 
\begin{table}[!ht]
    \centering
    \small
    \begin{tabular}{lrr}
    \hline\noalign{\smallskip}
    Count & DCIR & PMSI-MCO \\
    \noalign{\smallskip}\hline\noalign{\smallskip}
    Rows in the central table & 10,579,545,716 & 35,375,046\\
    Rows in the denormalized table & 10,636,094,654 & 3,208,682,967 \\
    Patients & 13,762,623 & 7,807,517\\
    Drug reimbursements events & 1,933,985,925 & NA \\
    Distinct drug codes & 16,289 & NA \\
    Reimbursed medical acts events & 210,847,422 & 97,484,303 \\
    Distinct medical acts codes & 7254 & 7591 \\
    Diagnoses events & NA & 120,212,253 \\
    Distinct diagnoses codes & NA & 16,895 \\
    Source data set disk size (CSV, GB) & 6,416.3 & 48.7\\
    Source data set disk size (Parquet, GB) & 572.7 & 5.9\\
    Flattened data set disk size (Parquet, GB) & 690.6 & 8.9 \\
    \hline\noalign{\smallskip}
    \end{tabular}
    \caption{Characteristics of the dataset used for experiments. Results are produced on a subset of SNDS containing 13.7 million subjects, followed up to three years. The scope is restricted to outpatient data (DCIR) and inpatient data excepted hospitalization at home, rehabilitation centers and psychiatric hospitals (PMSI-MCO). The central fact table of DCIR records cash flows resulting from healthcare reimbursements to patients covered by the French national healthcare insurance. One line in this table correspond to one cash flow (such as the reimbursement of a drug bought following a prescription). The central fact table of PMSI-MCO records hospital stays. 
    Events occurring during the stay are stored in dimension tables linked to this central table.}
    \label{tab:dataset}
\end{table}
Data from this sample is structured data containing common data types (timestamps, integers, floats, small strings), normalized according to the SNDS data model.
The testing data consisted in outpatient data (DCIR) and inpatient data excepted home hospitalization, rehabilitation centers and psychiatric hospitals (PMSI-MCO).
Raw data was extracted from the SNDS by CNAM, the French agency that manages this database. 
Extracts were dumped on the testing cluster as a set of CSV files.

SCALPEL3 was tested on a Mesos~\cite{hindman2011mesos} cluster of commodity servers with 14 worker nodes driven by 4 master nodes. 
Worker nodes resources amount to 224 2.4Ghz logical cores, 1.7Tb of RAM, and 448Tb of storage distributed over 88 spinning hard drives. 
These resources are shared over the cluster by HDFS~\cite{Shvachko:2010:HDF} for data storage and by Spark for memory storage and computations. 
This cluster and the configuration of the jobs were not fine-tuned for the usage of SCALPEL3, but follow standard guidelines for cluster configuration for distributed computing with Spark.

Denormalizing this dataset using SCALPEL-Flattening took about 6 hours using the 14 worker nodes. During the conversion of CSV tables to parquet files, worker nodes CPU and memory usage are maxed out on most worker nodes. During the join operations, resource usage is first dominated by network I/O to shuffle the data across the workers, followed by an increase in CPU and memory usage reaching two-thirds of the cluster capacity. 
Note that the current framework used for SNDS data cannot handle such denormalization so that there is no element of comparison for SCALPEL-Flattening with it.

SCALPEL-Extraction was evaluated on the following extraction tasks, that correspond to typical events required for public health research studying relations between fractures and some drug exposures:
(a) extraction of patient demographics (gender, age, eventual date of death), (b) extraction of drug dispenses, (c) filtering of patients w.r.t their first date of drug use (prevalent drug users, 65 drugs), (d) computation of drug exposures based on drug dispenses dates, (e) extraction of reimbursed medical acts, (f) extraction of diagnoses, (g) identification of fractures using the algorithm described in~\cite{bouyer_burden_2020} based on medical acts and diagnoses.

Indicative baseline performance was established by executing similar queries on the current SNDS infrastructure, based on SAS Enterprise Guide for analytics~\cite{sas}, connected to an Oracle SQL database hosted on Oracle Exadata servers~\cite{exadata}. 
This baseline performance was computed with a single run, as the current SNDS framework is designed to allocate resources dynamically each time a new query is submitted. 
The monitoring of resource usage on this SAS-Oracle infrastructure is not straightforward, since computations are divided between SAS and Oracle jobs, and since the resources of the Oracle Exadata infrastructure are divided across servers focused on storage or computation. 
At peak use (for task (c)), the Oracle job was using 10 CPUs supported by 4.9GB of PGA memory, while SAS was using 1 to 6GB of RAM.

\begin{figure}[!htp]
    \centering
    \includegraphics[width=0.87\textwidth]{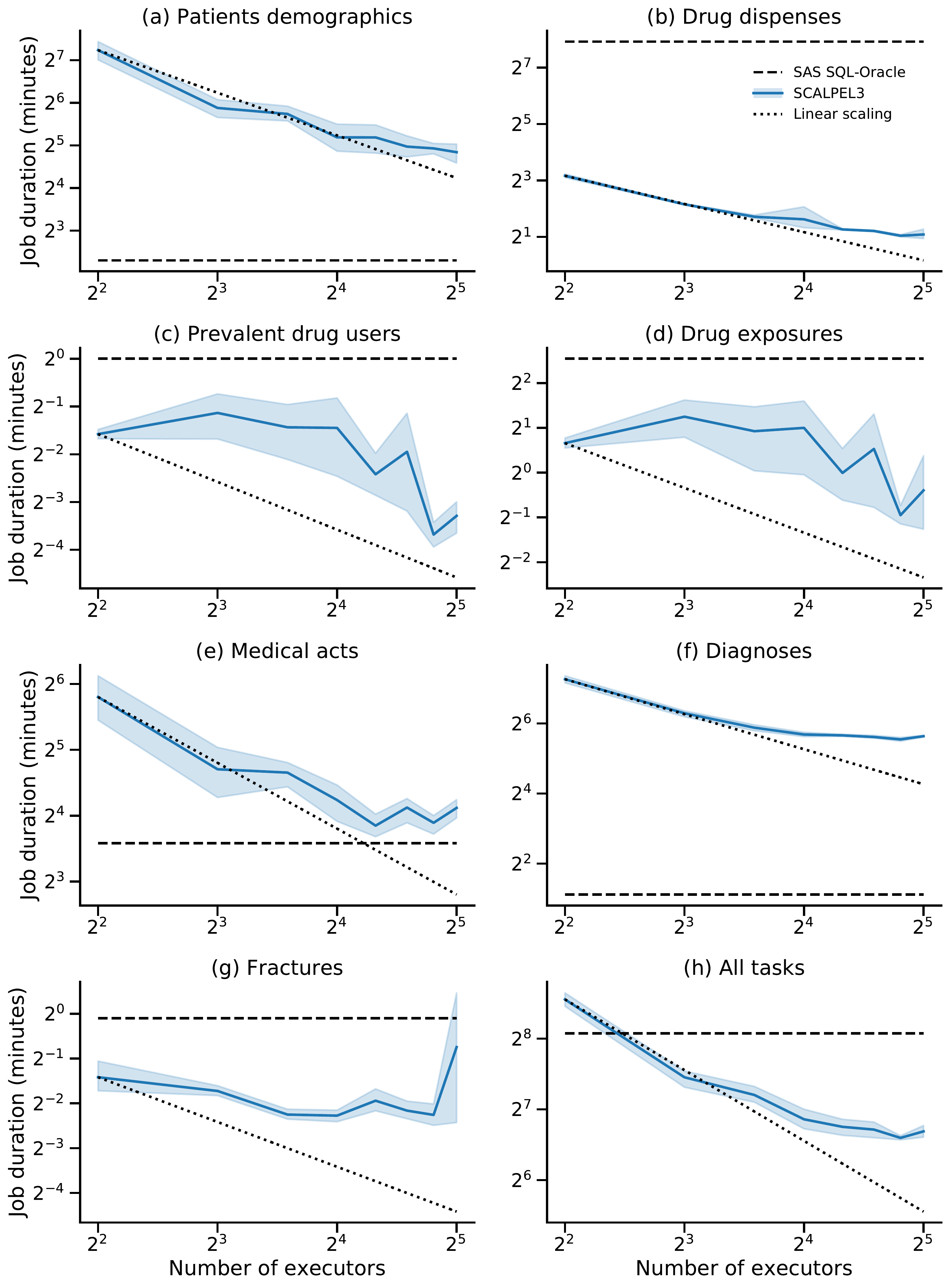}
    \caption{SCALPEL-Extraction scaling experiments. The blue solid line represents the mean total running time (in seconds) of queries (a)--(g) described in Section~\ref{sec:results} when varying the number of worker nodes used to perform the computation. Figure (h) represents the total running time of the (a)--(h) queries.
    Light blue bands represent one standard deviation computed over 5 runs. The dotted line corresponds to a theoretical performance assuming a perfect horizontal linear scaling (based on the single node performance). Dashed lines represent the runtime of similar queries on the SNDS SAS-Oracle infrastructure using a single run. Multiple runs were not performed on SAS-Oracle as computing resources are dynamically allocated for each queries and cannot be set beforehand.}
    \label{fig:scaling_benchmark}
\end{figure}

An assessment of the horizontal scaling of SCALPEL3 is performed by varying the number of executors (4~logical cores and 25~GB RAM) to perform these queries. 
All the results are displayed in Figure~\ref{fig:scaling_benchmark}.

SCALPEL-Analysis aims at providing useful abstractions to ease cohort data manipulation.
We provide in Supplementary Material, see Section~\ref{supp:examples} herein, examples that illustrate how these abstractions can be leveraged to perform typical data preparation in a few lines of code.

\section{Discussion}
\label{sec:discussion}

SCALPEL-Extraction reaches performances similar to SQL-SAS based SNDS framework when using 6 executors (Figure~\ref{fig:scaling_benchmark}(h)).
It is consistently faster on tasks involving large data volumes or complex operations such as tasks (b), (c), (d), and (g).
On the other hand, tasks involving the PMSI-MCO database (tasks (e) and (f)) exhibit poor performance. This is rooted in the flat table structure as PMSI-MCO is not sparse-by-block like DCIR (see the difference in the ratio of Rows in central table w.r.t. denormalized table in Table~\ref{tab:dataset}). It results in performing more tests on row values and data shuffle than necessary when performing queries on PMSI-MCO. Performance on these tasks could be further improved by slightly modifying the join strategy in the flattening step to ensure PMSI-MCO sparsity by block.

The cost of data denormalization should be considered to be fixed as this operation is done once and for all. The denormalized data can then be updated incrementally when new data are fed into the cluster (typically a few times a year).

SCALPEL-Extraction scales almost linearly from 4 to 16 executors. The scaling gains then slow down, reaching peak performance at 28 executors (see Figure~\ref{fig:scaling_benchmark}). These diminishing returns can be caused by the cluster resource sharing between storage services (HDFS) and computation (SCALPEL3). As a result, SCALPEL3 resource usage can be in conflict with HDFS resources as soon as the number of nodes used by SCALPEL3 excess one-third of the cluster\footnote{HDFS is configured to replicate the data across the worker nodes three times; HDFS performance is thus not much impacted if one-third of the nodes are not available at some point.}. Splitting the cluster nodes between storage nodes and computation nodes could improve horizontal scalability.
Note that for very small tasks (such as (c), (d), (g)), runtime is dominated by I/O operations and do not benefit particularly from additional CPUs.

Besides performance considerations, note that SCALPEL3 uses only open-source, free software and runs on commodity hardware, which is likely cheaper than Oracle Exadata servers and easier to scale if the data volume increases: a Spark cluster easily scales ``horizontally'' by adding more nodes.

The performance comparison between the two infrastructures is limited by (i) the impossibility to set the resources used by SAS-Oracle beforehand for these experiments does not allow for multiple runs and (ii) slight differences in query implementation caused by design differences such as columnar vs row orientation. Nonetheless, it shows that SCALPEL3 can be used as a viable open-source alternative running on commodity hardware while benefiting from horizontal scaling on very large jobs.

Besides, SCALPEL3 greatly improves the maintainability, audit, and reproducibility of studies using SNDS.
First, continuous integration of code updates and large code coverage (94\%) with unit testing is a big improvement in terms of maintainability over copy-pasted SQL snippets.
Secondly, SNDS expertise encapsulation for events extraction is fully tested and maintained in SCALPEL3, so it eases extraction algorithms reuse for studies and lowers the entry-barrier to SNDS.
Obviously, design and maintenance of SNDS concept extractors by a team of developers and SNDS specialists is a mandatory task, as the database contents are constantly evolving. Moreover, the relevance of extracted data (to answer a trade issue) requires some SNDS knowledge and is the responsibility of the user.

The combination of expert knowledge encapsulation (SCALPEL-Extraction) and 
interactive cohort manipulation (SCALPEL-Analysis) results in smaller and more readable user-code, leading to easily shared and reproducible studies, supported by data tracking and automated audit reports.
Finally, SCALPEL3 allows producing datasets compatible with several Python machine learning libraries formats, fostering methodological research on SNDS data, which was not possible with the proprietary software that is currently used.

The choice of the Python language might help SCALPEL3 adoption among the data science and machine learning community, while it might hinder its use among public health researchers who are traditionally using proprietary statistical softwares or the R language.
SCALPEL3 can be used in standalone mode\footnote{Using a single large server.} or in distributed mode\footnote{Using a computing cluster.} when working on large datasets. The knowledge and skills required to manage a computing cluster are not yet widespread which could also impede a large adoption of the distributed mode among small organizations.

Finally, while SCALPEL3 does not support international data standards yet, the development of vocabulary mapping tables in France was anticipated so as to ease future support of data standards such as OMOP-CDM~\cite{reisinger2010development} or FHIR~\cite{Bender:2013} to SCALPEL3.

\section{Conclusion}
SCALPEL3 could be further improved by optimizing the flattening step, so as to ensure optimal block-sparsity of the resulting denormalized databases automatically.
Besides, optimizing the cluster design to separate storage from computation as well as using YARN instead of Mesos to manage resources could help to improve its performance further by lowering data access times.
Finally, using Apache ORC~\cite{apache2018apache} instead of Parquet could also lead to further performance improvements. Parquet was initially chosen over ORC because of better integration with Spark. ORC is now well-integrated in it and has been reported to have better performances and a higher compression factor on non-nested data.

\section{Acknowledgments}
We thank the engineers who worked on this project at some point:  Firas Ben Sassi, Prosper Burq,  Philip Deegan, Daniel De Paula e Silva, Angel Francisco Orta, Xristos Giatsidis, Sathiya Prabhu Kumar. 

We also thank the people from CNAM or Polytechnique who were or are currently involved in the Polytechnique-CNAM partnership, namely, for CNAM : Muhammad Abdallah, Aurélie Bannay, Hélène Caillol, Anthony Du, Sébastien Dumontier, Mehdi Gabbas, Claude Gissot, Moussa Laanani, Mickaël Lechapelier, Clémence Martin, Anke Neumann, Cédric Pulrulczyk, Jérémie Rudant, Omar Sow, Kévin Vu Saintonge, Alain Weill, and for Polytechnique: Qing Chen, Agathe Guilloux, Anastasiia Nitavskyi, Yiyang Yu.


\newpage
\appendix
\section*{Supplementary material}

\section{Scalpel Analysis usage examples}
\label{supp:examples}

This section presents a quick example of the SCALPEL-Analysis API. {\color{incolor}\texttt{In [ ]}} and {\color{outcolor}\texttt{Out [ ]}} respectively indicate numbered input code and output results.
Example \texttt{[1]} shows how to load a cohort collection from a json file produced by SCALPEL-Extraction. Examples \texttt{[2]}, \texttt{[3]}, \texttt{[4]} show how to access a cohort from a cohort collection and to count their subjects.
Example \texttt{[5]} shows how to use algebraic manipulations over cohort to remove prevalent cases from a given population, and times this operation to show that it fast enough for interactive use.
Example \texttt{[6]} highlight automatically generated captions for cohorts resulting from algebraic operations, while examples \texttt{[7]} and \texttt{[8]} show how to access subject and event data from a cohort.
Examples \texttt{[9]} and \texttt{[10]} illustrate how to use SCALPEL-Analysis to define a \texttt{CohortFlow} from a sequence of cohorts, then using \texttt{scalpel.stats} to obtain statistics about the distributions of gender and age along the stages. In example \texttt{[9]}, excluding patients with a fracture does not introduce much changes in the gender and age distributions. In example \texttt{[10]} however, keeping only patients with fractures in the final stage leads to an older population, with an important change in the age distribution of women (a well-known phenomenon related to osteoporosis).

{\small
\begin{Verbatim}[commandchars=\\\{\}]
{\color{incolor}In [{\color{incolor}1}]:} \PY{k+kn}{from} \PY{n+nn}{scalpel}\PY{n+nn}{.}\PY{n+nn}{core}\PY{n+nn}{.}\PY{n+nn}{cohort\PYZus{}collection} \PY{k}{import} \PY{n}{CohortCollection}
        
        \PY{c+c1}{\PYZsh{} metadata\PYZus{}path = \PYZsq{}/path/to/some/metadata\PYZus{}file.json\PYZsq{}}
        \PY{n}{cc} \PY{o}{=} \PY{n}{CohortCollection}\PY{o}{.}\PY{n}{from\PYZus{}json}\PY{p}{(}\PY{n}{metadata\PYZus{}path}\PY{p}{)}
        \PY{n+nb}{print}\PY{p}{(}\PY{n}{cc}\PY{o}{.}\PY{n}{cohorts\PYZus{}names}\PY{p}{)}
\end{Verbatim}

\begin{Verbatim}[commandchars=\\\{\}]
{\color{outcolor}Out[{\color{outcolor}1}]:} \{'follow\_up', 'acts', 'fractures', 'extract\_hospital\_stays',
         'filter\_patients', 'liberal\_acts', 'extract\_patients', 'exposures',
         'diagnoses', 'drug\_purchases'\}

\end{Verbatim}

\begin{Verbatim}[commandchars=\\\{\}]
{\color{incolor}In [{\color{incolor}2}]:} \PY{n}{base\PYZus{}population} \PY{o}{=} \PY{n}{cc}\PY{o}{.}\PY{n}{get}\PY{p}{(}\PY{l+s+s1}{\PYZsq{}}\PY{l+s+s1}{extract\PYZus{}patients}\PY{l+s+s1}{\PYZsq{}}\PY{p}{)}
        \PY{n}{base\PYZus{}population}\PY{o}{.}\PY{n}{subjects}\PY{o}{.}\PY{n}{count}\PY{p}{(}\PY{p}{)}
\end{Verbatim}

\begin{Verbatim}[commandchars=\\\{\}]
{\color{outcolor}Out[{\color{outcolor}2}]:} 5186601
\end{Verbatim}
            
\begin{Verbatim}[commandchars=\\\{\}]
{\color{incolor}In [{\color{incolor}3}]:} \PY{n}{exposed\PYZus{}subjects} \PY{o}{=} \PY{n}{cc}\PY{o}{.}\PY{n}{get}\PY{p}{(}\PY{l+s+s1}{\PYZsq{}}\PY{l+s+s1}{exposures}\PY{l+s+s1}{\PYZsq{}}\PY{p}{)}
        \PY{n}{exposed\PYZus{}subjects}\PY{o}{.}\PY{n}{subjects}\PY{o}{.}\PY{n}{count}\PY{p}{(}\PY{p}{)}
\end{Verbatim}

\begin{Verbatim}[commandchars=\\\{\}]
{\color{outcolor}Out[{\color{outcolor}3}]:} 2666662
\end{Verbatim}
            
\begin{Verbatim}[commandchars=\\\{\}]
{\color{incolor}In [{\color{incolor}4}]:} \PY{n}{fractured\PYZus{}subjects} \PY{o}{=} \PY{n}{cc}\PY{o}{.}\PY{n}{get}\PY{p}{(}\PY{l+s+s1}{\PYZsq{}}\PY{l+s+s1}{fractures}\PY{l+s+s1}{\PYZsq{}}\PY{p}{)}
         \PY{n}{fractured\PYZus{}subjects}\PY{o}{.}\PY{n}{subjects}\PY{o}{.}\PY{n}{count}\PY{p}{(}\PY{p}{)}
\end{Verbatim}

\begin{Verbatim}[commandchars=\\\{\}]
{\color{outcolor}Out[{\color{outcolor}4}]:} 179072
\end{Verbatim}
            
\begin{Verbatim}[commandchars=\\\{\}]
{\color{incolor}In [{\color{incolor}5}]:} \PY{o}{\PYZpc{}\PYZpc{}}\PY{k}{timeit}
         \PY{c+c1}{\PYZsh{} Select subjects in base population who were exposed but }
         \PY{c+c1}{\PYZsh{} have not experienced a fracture}
         final\PYZus{}cohort = (exposed\PYZus{}subjects.intersection(base\PYZus{}population)
                             ).difference(fractured\PYZus{}subjects)
         final\PYZus{}cohort.subjects.count()
\end{Verbatim}

\begin{Verbatim}[commandchars=\\\{\}]
11.3 s ± 4.5 s per loop (mean ± std. dev. of 7 runs, 1 loop each)

\end{Verbatim}

\begin{Verbatim}[commandchars=\\\{\}]
{\color{outcolor}Out[{\color{outcolor}5}]:} 2542922
\end{Verbatim}
            
\begin{Verbatim}[commandchars=\\\{\}]
{\color{incolor}In [{\color{incolor}6}]:} \PY{n}{final\PYZus{}cohort}\PY{o}{.}\PY{n}{describe}\PY{p}{(}\PY{p}{)}
\end{Verbatim}

\begin{Verbatim}[commandchars=\\\{\}]
{\color{outcolor}Out[{\color{outcolor}6}]:} 'Events are exposures. Events contain only subjects
         with event exposures with extract\_patients without
         subjects with event fractures.'
\end{Verbatim}

\begin{Verbatim}[commandchars=\\\{\}]
{\color{incolor}In [{\color{incolor}7}]:} \PY{n}{final\PYZus{}cohort}\PY{o}{.}\PY{n}{subjects}\PY{o}{.}\PY{n}{show}\PY{p}{(}\PY{p}{)}
\end{Verbatim}

\begin{Verbatim}[commandchars=\\\{\}]
+---------+------+-------------------+-------------------+
|patientID|gender|          birthDate|          deathDate|
+---------+------+-------------------+-------------------+
|    Alice|     2|1934-07-27 00:00:00|               null|
|      Bob|     1|1951-05-01 00:00:00|               null|
|   Carole|     2|1942-01-12 00:00:00|               null|
|    Chuck|     1|1933-10-03 00:00:00|2011-06-20 00:00:00|
|    Craig|     1|1943-07-27 00:00:00|2012-12-10 00:00:00|
|      Dan|     1|1971-10-07 00:00:00|               null|
|     Erin|     2|1924-01-12 00:00:00|               null|
|      Eve|     2|1953-02-21 00:00:00|               null|
+---------+------+-------------------+-------------------+

\end{Verbatim}

\begin{Verbatim}[commandchars=\\\{\}]
{\color{incolor}In [{\color{incolor}8}]:} \PY{n}{final\PYZus{}cohort}\PY{o}{.}\PY{n}{events}\PY{o}{.}\PY{n}{show}\PY{p}{(}\PY{p}{)}
\end{Verbatim}

\begin{Verbatim}[commandchars=\\\{\}]
+---------+--------+-------+-----+------+-------------------+-------------------+
|patientID|category|groupID|value|weight|              start|                end|
+---------+--------+-------+-----+------+-------------------+-------------------+
|    Alice|exposure|   null|DrugA|   1.0|2013-08-08 00:00:00|2013-10-07 00:00:00|
|    Alice|exposure|   null|DrugB|   1.0|2012-09-11 00:00:00|2012-12-30 00:00:00|
|    Alice|exposure|   null|DrugC|   1.0|2013-01-23 00:00:00|2013-03-24 00:00:00|
|      Bob|exposure|   null|DrugB|   1.0|2014-03-04 00:00:00|2014-05-03 00:00:00|
|   Carole|exposure|   null|DrugB|   1.0|2010-01-25 00:00:00|2010-12-13 00:00:00|
|      Dan|exposure|   null|DrugA|   1.0|2012-11-29 00:00:00|2013-01-28 00:00:00|
|     Erin|exposure|   null|DrugC|   1.0|2010-09-09 00:00:00|2011-01-17 00:00:00|
|      Eve|exposure|   null|DrugA|   1.0|2010-04-30 00:00:00|2010-08-02 00:00:00|
+---------+--------+-------+-----+------+-------------------+-------------------+

\end{Verbatim}

\begin{Verbatim}[commandchars=\\\{\}]
{\color{incolor}In [{\color{incolor}9}]:} \PY{k+kn}{from} \PY{n+nn}{scalpel}\PY{n+nn}{.}\PY{n+nn}{stats}\PY{n+nn}{.}\PY{n+nn}{patients} \PY{k}{import} \PY{n}{distribution\PYZus{}by\PYZus{}gender\PYZus{}age\PYZus{}bucket}
         \PY{k+kn}{from} \PY{n+nn}{scalpel}\PY{n+nn}{.}\PY{n+nn}{core}\PY{n+nn}{.}\PY{n+nn}{cohort\PYZus{}flow} \PY{k}{import} \PY{n}{CohortFlow}
         
         \PY{n}{flow} \PY{o}{=} \PY{n}{CohortFlow}\PY{p}{(}\PY{p}{[}\PY{n}{base\PYZus{}population}\PY{p}{,} \PY{n}{exposed\PYZus{}subjects}\PY{p}{,} \PY{n}{final\PYZus{}cohort}\PY{p}{]}\PY{p}{)}
         
         \PY{k}{for} \PY{n}{cohort} \PY{o+ow}{in} \PY{n}{flow}\PY{o}{.}\PY{n}{steps}\PY{p}{:}
             \PY{n}{figure} \PY{o}{=} \PY{n}{plt}\PY{o}{.}\PY{n}{figure}\PY{p}{(}\PY{n}{figsize}\PY{o}{=}\PY{p}{(}\PY{l+m+mi}{8}\PY{p}{,} \PY{l+m+mf}{4.5}\PY{p}{)}\PY{p}{)}
             \PY{n}{distribution\PYZus{}by\PYZus{}gender\PYZus{}age\PYZus{}bucket}\PY{p}{(}\PY{n}{cohort}\PY{o}{=}\PY{n}{cohort}\PY{p}{,} \PY{n}{figure}\PY{o}{=}\PY{n}{figure}\PY{p}{)}
\end{Verbatim}


\begin{center}
\adjustimage{max size={0.7\linewidth}{0.9\paperheight}}{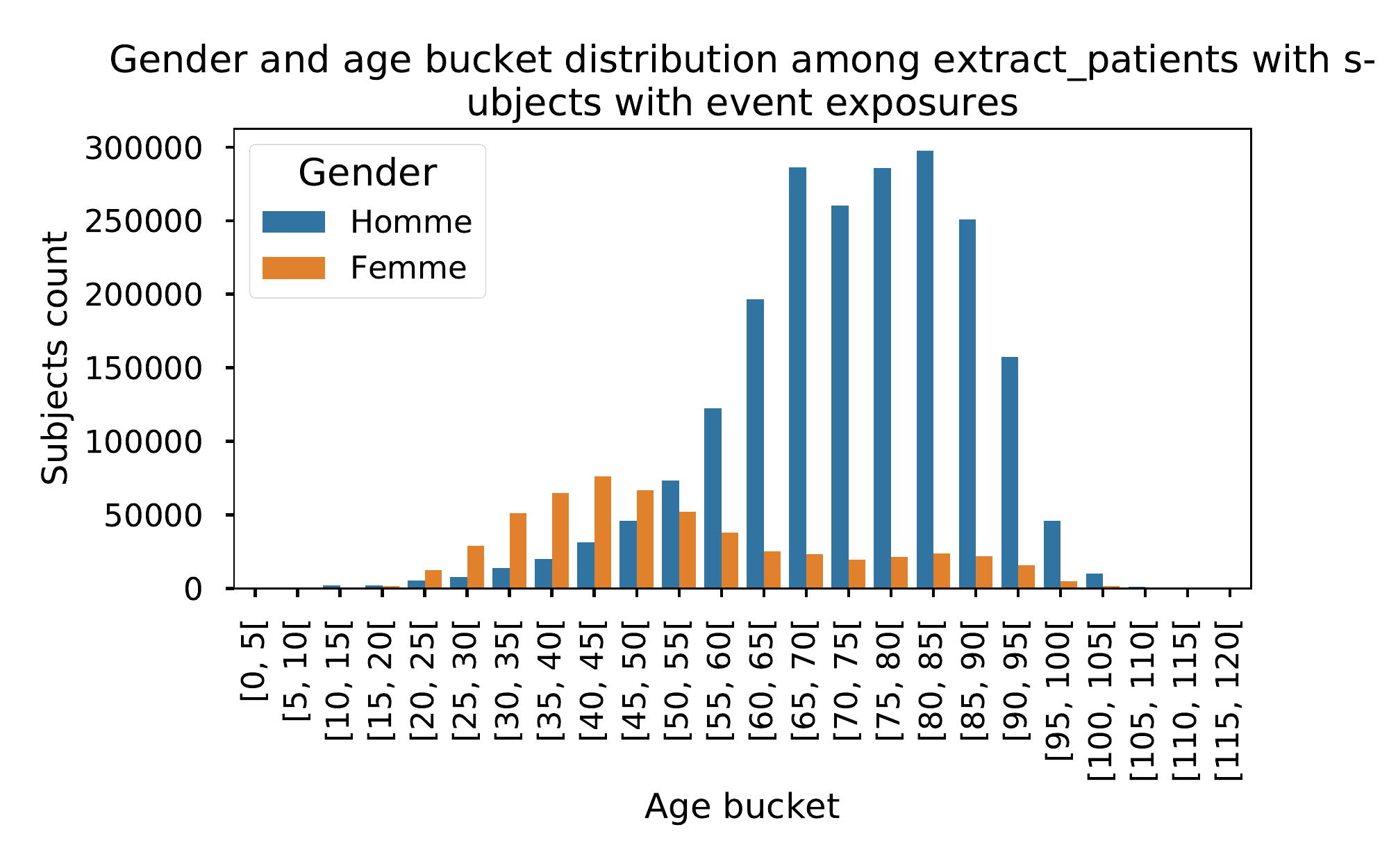}
\end{center}
{ \hspace*{\fill} \\}

\begin{center}
\adjustimage{max size={0.7\linewidth}{0.9\paperheight}}{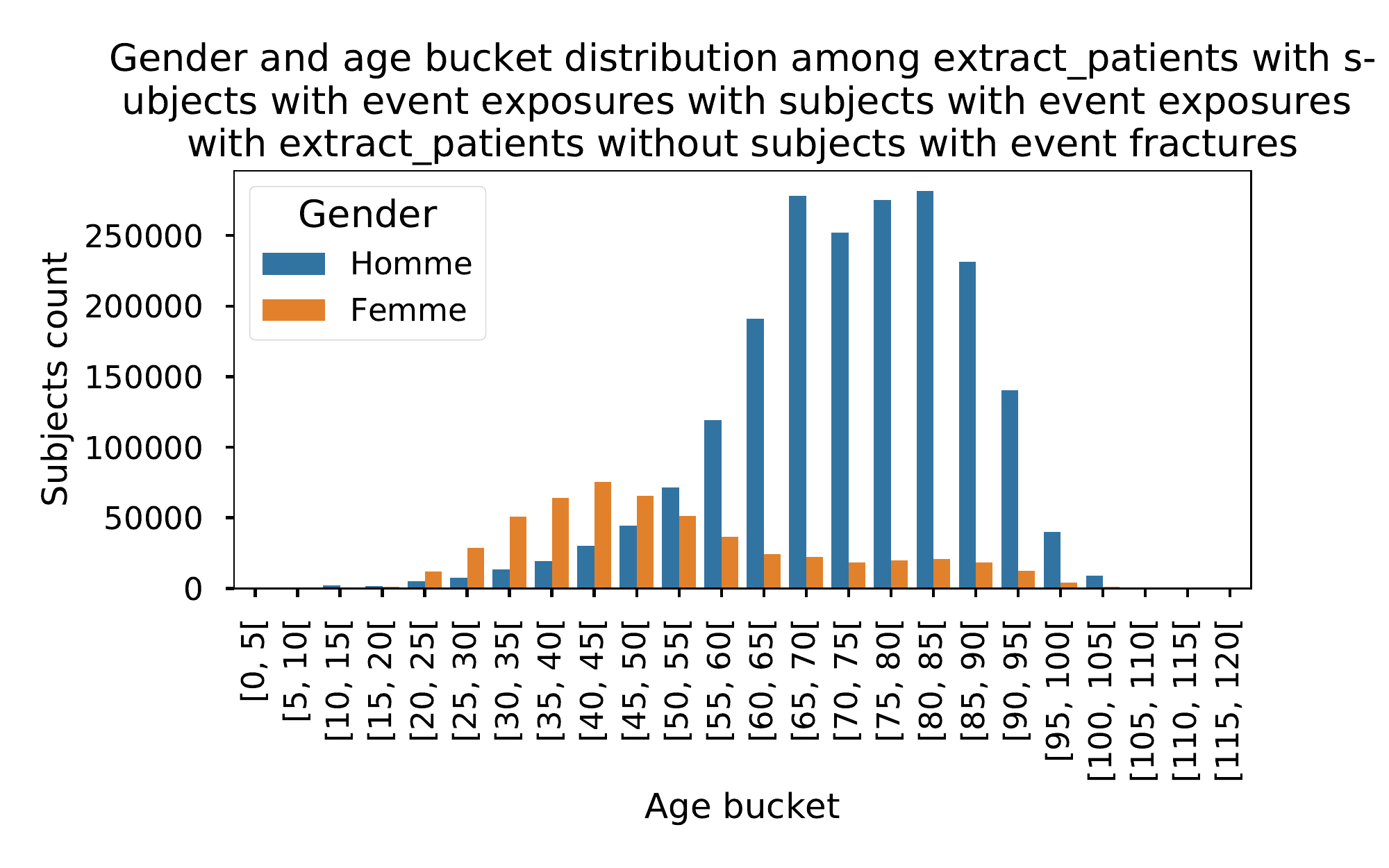}
\end{center}
{ \hspace*{\fill} \\}
    
\begin{Verbatim}[commandchars=\\\{\}]
{\color{incolor}In [{\color{incolor}10}]:} \PY{k+kn}{from} \PY{n+nn}{scalpel}\PY{n+nn}{.}\PY{n+nn}{stats}\PY{n+nn}{.}\PY{n+nn}{patients} \PY{k}{import} \PY{n}{distribution\PYZus{}by\PYZus{}gender\PYZus{}age\PYZus{}bucket}
         \PY{k+kn}{from} \PY{n+nn}{scalpel}\PY{n+nn}{.}\PY{n+nn}{core}\PY{n+nn}{.}\PY{n+nn}{cohort\PYZus{}flow} \PY{k}{import} \PY{n}{CohortFlow}
         
         \PY{n}{flow} \PY{o}{=} \PY{n}{CohortFlow}\PY{p}{(}\PY{p}{[}\PY{n}{base\PYZus{}population}\PY{p}{,} \PY{n}{exposed\PYZus{}subjects}\PY{p}{,} \PY{n}{fractured\PYZus{}subjects}\PY{p}{]}\PY{p}{)}
         
         \PY{k}{for} \PY{n}{cohort} \PY{o+ow}{in} \PY{n}{flow}\PY{o}{.}\PY{n}{steps}\PY{p}{:}
             \PY{n}{figure} \PY{o}{=} \PY{n}{plt}\PY{o}{.}\PY{n}{figure}\PY{p}{(}\PY{n}{figsize}\PY{o}{=}\PY{p}{(}\PY{l+m+mi}{8}\PY{p}{,} \PY{l+m+mf}{4.5}\PY{p}{)}\PY{p}{)}
             \PY{n}{distribution\PYZus{}by\PYZus{}gender\PYZus{}age\PYZus{}bucket}\PY{p}{(}\PY{n}{cohort}\PY{o}{=}\PY{n}{cohort}\PY{p}{,} \PY{n}{figure}\PY{o}{=}\PY{n}{figure}\PY{p}{)}
\end{Verbatim}


\begin{center}
\adjustimage{max size={0.7\linewidth}{0.9\paperheight}}{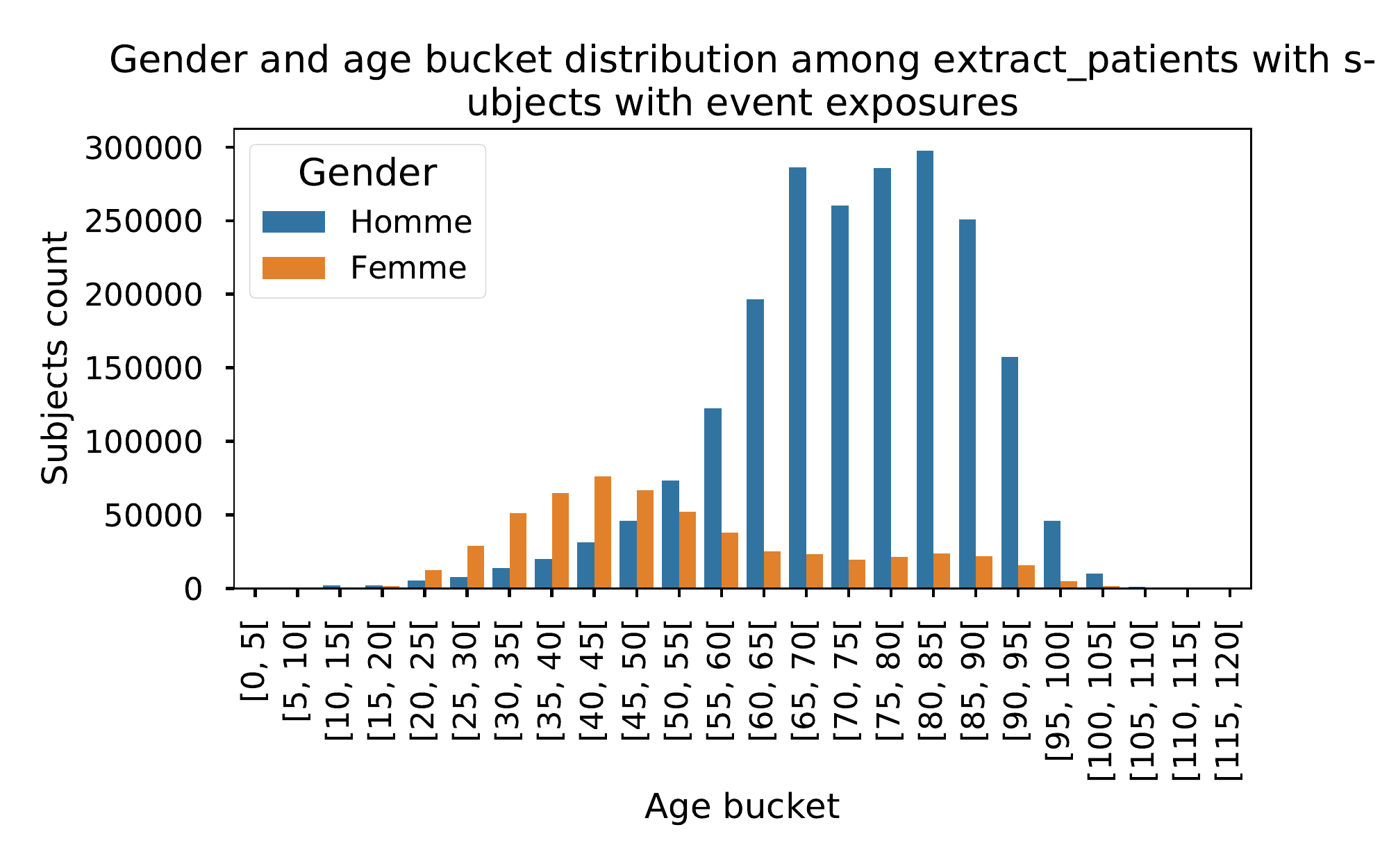}
\end{center}
{ \hspace*{\fill} \\}

\begin{center}
\adjustimage{max size={0.7\linewidth}{0.9\paperheight}}{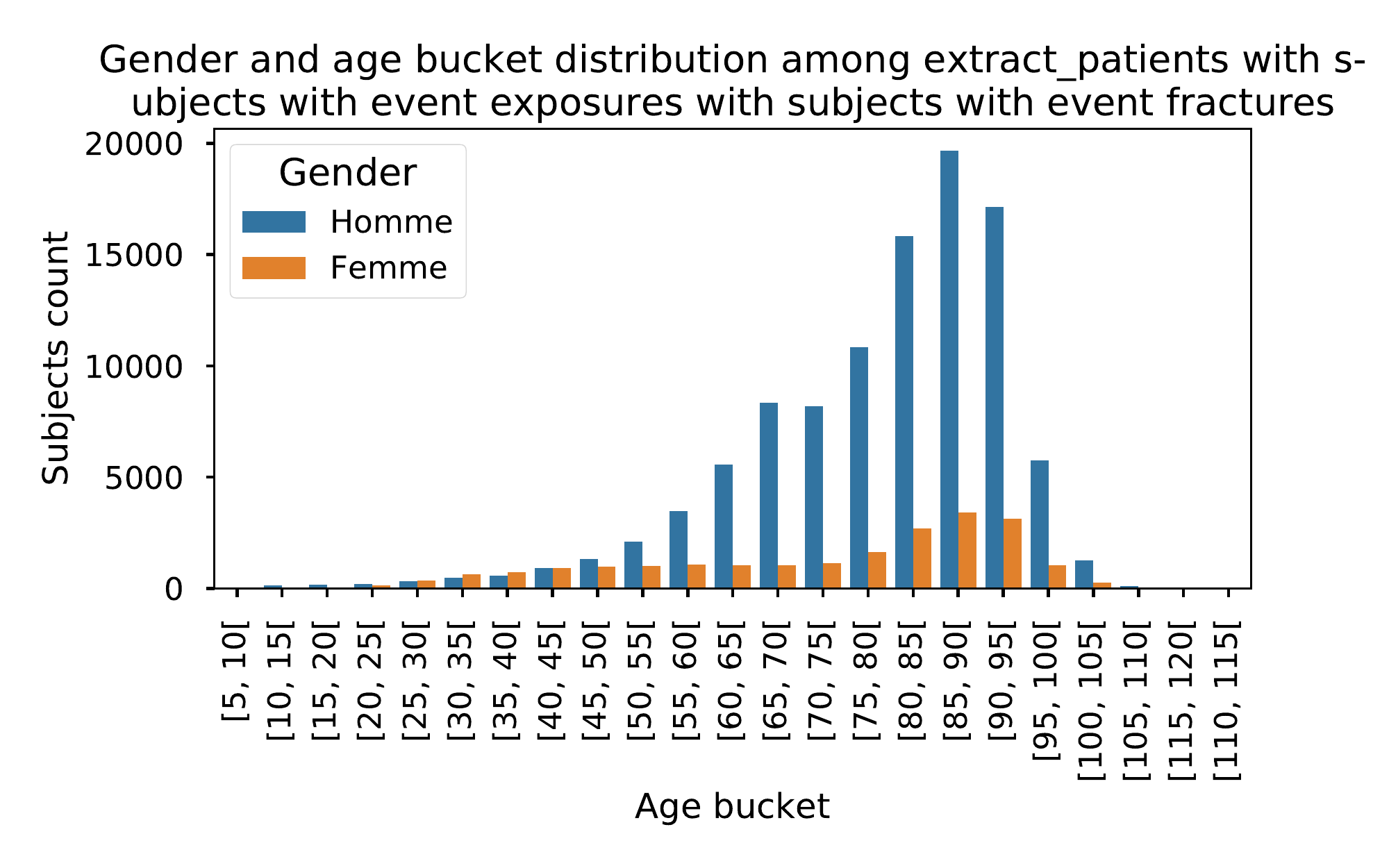}
\end{center}
{ \hspace*{\fill} \\}
}

\newpage
\section{List of SNDS databases currently denormalized.}

\begin{table}[!h]
    \centering
    \begin{tabular}{ll}
    \hline\noalign{\smallskip}
    Database & Contents \\
    \hline\noalign{\smallskip}
    DCIR & Outpatients reimbursement data \\
    \emph{PMSI} & Hospital discharges \\
    $\qquad$MCO & Acute ward \\
    $\qquad$MCO CE & Acute ward outpatients treatment \\
    $\qquad$SSR & Rehabilitation \\
    $\qquad$SSR CE & Rehabilitation outpatients treatment \\
    $\qquad$HAD & Home-to-home care \\
    IR\_IMB\_R & Long term chronic diseases \\
    IR\_BEN\_R & Patients socio-demographic information \\
    \hline\noalign{\smallskip}
    \end{tabular}
    \caption{List of SNDS sub-databases which are currently denormalized by SCALPEL-Flattening. IR\_IMB\_R and IR\_BEN\_R are tables and were simply converted to Parquet files.}
    \label{tab:flattened_databases}
\end{table}

\newpage
\section{List of available extractors}

\begin{table}[!h]
    \centering
    \small
    \begin{tabular}{llp{11em}l}
    \hline\noalign{\smallskip}
    Extractor & & Source databases & Event Type \\
    \hline\noalign{\smallskip}
    \emph{Medical acts} & & & \\
    $\qquad$CCAM & & DCIR, MCO, MCOCE, SSR, SSRCE, HAD & Punctual \\
    $\qquad$NGAP & & DCIR, MCOCE & Punctual \\
    $\qquad$CSARR & & SSR & Punctual \\
    Biological acts & & DCIR & Punctual \\
    \emph{Practitioner encounter} & & & \\
    $\qquad$Medical & & DCIR & Punctual \\
    $\qquad$Non-medical & & DCIR & Punctual \\
    Drug dispenses & & DCIR & Punctual \\
    \emph{Diagnoses} & & & \\
    $\qquad$Main & & MCO, SSR, HAD & Punctual \\
    $\qquad$Associated & & MCO, SSR, HAD & Punctual \\
    $\qquad$Linked & & MCO, SSR, HAD & Punctual \\
    $\qquad$Long-term chronic disease & & IR\_IMB\_R & Longitudinal \\
    Hospital stay & & MCO & Longitudinal \\
    Emergency visit & & MCOCE & Punctual \\
    SSR Stay & & SSR & Longitudinal \\
    \emph{Hospital takeover} & & SSR, HAD & Punctual \\
    $\qquad$Main Takeover reason & & HAD & Punctual \\
    $\qquad$Associated Takeover reason & & HAD & Punctual \\
    Patient & & IR\_BEN\_R, DCIR, MCO, SSR, HAD & Person \\
    \hline\noalign{\smallskip}
    \end{tabular} 
    \caption{List of implemented event extractors. This list is meant to grow over time. More details are available in SCALPEL-Extraction~\cite{SCALPEL-Extraction} wiki on GitHub at \url{https://github.com/X-DataInitiative/SCALPEL-Extraction/wiki}.}
    \label{tab:extractors}
\end{table}

\newpage
\section{List of the available transformers}

\begin{table}[!h]
    \centering
    \small
    \begin{tabular}{ll}
    \hline\noalign{\smallskip}
    Transformer & Source events [optional] \\
    \hline\noalign{\smallskip}
    Observation period & Patients, [Any] \\
    Trackloss & Patients, [drug dispenses] \\
    Follow-up & Patients, observation period, [trackloss, drug dispenses, diagnoses] \\
    Drug prescription & Drug dispenses \\
    Drug interaction & Drug dispenses \\
    \emph{Exposure} & \\
    $\qquad$Limited in time & Drug dispenses, Follow-up, [drug interaction]\\
    $\qquad$Unlimited & Drug dispenses, Follow-up, [drug interaction] \\
    \emph{Outcomes} & \\
    $\qquad$Fractures per body site & Medical acts, diagnoses \\
    $\qquad$Bladder cancer & Medical acts, diagnoses \\
    $\qquad$Infarctus & Diagnoses \\
    $\qquad$Heart failure & Diagnoses \\
    \hline\noalign{\smallskip}
    \end{tabular}
    \caption{List of implemented transformers. This list is meant to grow over time. More details are available in SCALPEL-Extraction~\cite{SCALPEL-Extraction} wiki on GitHub at \url{https://github.com/X-DataInitiative/SCALPEL-Extraction/wiki}.}
    \label{tab:transformers}
\end{table}

\end{document}